\documentclass{emulateapj}
\usepackage{graphicx}
\begin{document}

\newcommand{\arcm}{$^\prime$}
\newcommand{\arcs}{$^{\prime\prime}$}
\newcommand{\m}{$^{\rm m}\!\!.$}
\newcommand{\D}{$^{\rm d}\!\!.$}
\newcommand{\F}{$^{\rm P}\!\!.$}
\newcommand{\kms}{km~s$^{-1}$}
\newcommand{\ks}{km~s$^{-1}$}
\newcommand{\ms}{M$_{\odot}$}
\newcommand{\rs}{R$_{\odot}$}

\newcommand{\hip}{$Hipparcos$}
\newcommand{\ond}{Ond\v{r}ejov}
\newcommand{\ova}{Ostrava}
\newcommand{\valmez}{Vala\v{s}sk\'e Mezi\v{r}\'{\i}\v{c}\'{\i}}
\newcommand{\jil}{J\'{\i}lov\'e}
\newcommand{\hra}{Hradec Kr\'alov\'e}
\newcommand{\pec}{Pec pod Sn\v{e}\v{z}kou}
\newcommand{\La}{La~Silla}

\title{\textbf{Apsidal motion and absolute parameters \\ for five LMC eccentric eclipsing binaries}
 \thanks{Based on observations made with ESO Telescopes at the La Silla Paranal Observatory under programme ID
 68.A-0223(A), and on data collected with the Danish 1.54 m telescope at the ESO La Silla Observatory.
        Tables A.1 and B.1 are only available in electronic form at
       {\tt http://www.aanda.org} } }

\author{P. Zasche~\altaffilmark{1}
    \and M. Wolf~\altaffilmark{1}
            }

\affil{Astronomical Institute, Charles University in Prague, Faculty of Mathematics and Physics, CZ-180~00 Praha 8, \\
             V~Hole\v{s}ovi\v{c}k\'ach 2, Czech Republic, email: zasche@sirrah.troja.mff.cuni.cz}

\date{Received \today}


\begin{abstract}
 {Aims: As part of our observational projects at the La Silla Danish 1.54-meter telescope, we aim to
measure the precise times of minimum light for eccentric eclipsing binaries in the Large Magellanic
Cloud, needed for accurate determination of apsidal motion. Many new times of minima were derived
from the photometric databases OGLE and MACHO. Several new minima were also observed. Five
early-type and eccentric-orbit eclipsing binaries:
        HV~982  ($P=5\fd34$, $e=0.15$),
        HV~2274 ($5\fd73, 0.17$),
        MACHO~78.6097.13 ($3\fd11, 0.05$),
        MACHO~81.8881.47 ($3\fd88, 0.22$), and
        MACHO~79.5377.76 ($2\fd64, 0.06$) were studied. }\\[-3mm]

 {Methods: The $O-C$\ diagrams of the systems were analysed using all reliable timings
found in the literature, and new or improved elements of apsidal motion were obtained. Light and
radial velocity curves of MACHO~81.8881.47 and MACHO~79.5377.76 were analysed using the program
{\sc PHOEBE}.}\\[-3mm]

 {Results: We derived for the first time or significantly improved the relatively
short periods of apsidal motion of 211 (12), 127 (8), 48 (13), 103 (20), and 42 (19) years,
respectively. The internal structure constants, log $k_2$, were found to be -2.37, -2.47, -2.17,
-2.02, and -1.86 respectively, under the assumption that the component stars rotate
pseudosynchronously. The relativistic effects are weak, up to 6\% of the total apsidal motion rate.
The masses for MACHO~81.8881.47 resulted in 5.51 (0.21) and 5.40 (0.19) $M_\odot$, while for
MACHO~79.5377.76 the masses are 11.26 (0.35) and 11.27 (0.35) $M_\odot$, respectively.}
 \end{abstract}



\section{Introduction}

Eccentric eclipsing binaries (EEB) with an apsidal motion can provide us with an important
observational test of theoretical models of stellar structure and evolution. A long-term collecting
the times of minima of EEBs observed throughout the apsidal motion cycle and consecutive detailed
analysis of the period variations of EEB can be performed, yielding both the orbital eccentricity
and the period of rotation of the apsidal line with high accuracy (Gim\'enez 1994). Many different
sets of stellar evolution models have been published in recent years, e.g. Maeder (1999), Claret
(2004), or Claret (2006); however, to distinguish between them and to test which one is more
suitable is still rather difficult. The internal structure constants as derived from the apsidal
motion analysis could serve as one independent criterion. On the other hand, to discriminate
between the models only stellar parameters for EEBs with an accuracy of 1\% can be used.

The Magellanic Clouds are of prime importance in the context of stellar evolution theory. However,
the chemical composition of the Magellanic Clouds differs from that of the solar neighborhood (e.g.
Ribas 2004) and the study of these massive and metal-deficient stars in the LMC checks our
evolutionary models for these abundances. All eclipsing binaries analysed here have properties that
make them important astrophysical laboratories for studying the structure and evolution of massive
stars (Ribas 2004).

Here we analyse the observational data and rates of apsidal motion for five LMC detached eclipsing
systems. All these systems are early-type objects known to have eccentric orbits and to exhibit
apsidal motion. Similar studies of LMC EEBs have been presented by Michalska \& Pigulski (2005,
hereafter MP05) and Michalska (2007).

\section{Observations of minimum light}

The monitoring of faint EEBs in external galaxies requires only moderate telescopes in the 1 - 2m
class range equipped with a modern CCD camera. However, a large amount of observing time is needed,
which is unavailable at larger telescopes. During past observational seasons, we accumulated over
1600 photometric observations at selected phases during primary and secondary eclipses and derived
16 precise times of minimum light for selected eccentric systems. New CCD photometry was obtained
at the La Silla Observatory in Chile, where the 1.54 m Danish telescope (hereafter DK154) with the
CCD camera and \textit{RI} filters was used.

All CCD measurements were dark-subtracted and then flat-fielded using sky exposures taken at either
dusk or dawn. Several comparison stars were chosen in the same frame as the variables. A synthetic
aperture photometry and astrometry software developed by M.~Velen and P.~Pravec called {\sc Aphot},
was routinely used for data obtained. No correction for differential extinction was applied because
of the proximity of the comparison stars to the variable and the resulting negligible differences
in air mass and their similar spectral types.

The new times of primary and secondary minima and their errors were determined by the classical
Kwee-van Woerden (1956) algorithm. All new times of minima are given in Table~A.1 in the Appendix,
where epochs are calculated from the ephemeris given in Table~\ref{t2}; the other columns are
self-evident.

\section{Photometry}

For all of the systems we harvested the {\sc Macho} (Faccioli et al. 2007) and {\sc Ogle} (Graczyk
et al. 2011) photometry available online. These photometric data were used both for minima times
analysis as well as for light curve analysis.

The analysis of the light curves for two of the systems was carried out using the program {\sc
PHOEBE}, ver. 0.31a (Pr{\v s}a \& Zwitter 2005), which is based on the Wilson-Devinney algorithm
(Wilson \& Devinney 1971) and its later modifications, but some of the parameters had to be fixed
during the fitting process. The albedo coefficient remained fixed at value 1.0 and the gravity
darkening coefficients $g = 1.0$. The limb darkening coefficients were interpolated from the van
Hamme's tables (van Hamme 1993). A problem emerged with the synchronicity parameters ($F_i$) due to
poor coverage of the RV data near the eclipses and low quality of the light curves used for the LC
analysis, hence we fixed these values at $F_i = 0$. The temperature of the primary component was
derived from the $V-I$ photometric index, from MP05 and from the resulting masses derived from the
combined LC+RV analysis.

\section{Spectroscopy}

The spectroscopic data for two of the systems (MACHO~81.8881.47, MACHO~79.5377.76) were found in
the ESO Archive of the UV-Visual Echelle Spectrograph (UVES) at the Very Large Telescope (VLT). The
spectra were obtained during the ESO Period 68 program ``\emph{Precise distances to the LMC and SMC
from double-lined eclipsing binaries}"; the PI of the project was A. Clausen. Typical exposure
times were about 1500 seconds, while the spectra typically have a signal-to-noise ($S/N$) ratio of
about 50. The original data were reduced using the standard ESO routines. The final radial
velocities (hereafter RV) used for the analysis were derived via a manual cross-correlation
technique (i.e. the direct and flipped profile of spectral lines manually shifted on the computer
screen to achieve the best match) using program SPEFO (Horn et al. 1996, \v{S}koda 1996) on several
absorbtion lines in the measured spectral region (usually $H_\beta$ to $H_\vartheta$). The derived
radial velocities are given in Table \ref{RVs}. We estimate the error of individual data points to
be about 5 km/s.

\section{An approach for the analysis}

For the analysis we used the approach presented below. For the systems where the inclination of the
orbit is known, the first two steps can be skipped.
 \begin{itemize}
   \item First, all of the available photometric data were analysed, resulting in a set
   of minima times. Preliminary apsidal motion parameters were derived (with the assumption $i=90^\circ$).\\[-1mm]

   \item Second, the eccentricity ($e$), argument of periastron ($\omega$), and apsidal motion
   rate ($\dot \omega$) derived from the apsidal motion analysis were used for the preliminary
   light curve (hereafter LC) analysis.\\[-1mm]

   \item Third, the inclination ($i$) from the LC analysis was used for the final apsidal motion
   analysis.\\[-1mm]

   \item And finally, the resulting $e$, $\omega$, and $\dot \omega$ values from the apsidal motion
   analysis were used for the final LC + RV analysis.\\[-1mm]
 \end{itemize}

\section{Apsidal motion analysis}

The apsidal motion in all eccentric systems was studied by means of an $O-C$\ diagram analysis. All
available times of minima, both published in the literature and newly measured, were analysed using
the method presented by Gim\'enez \& Garc\'{\i}a-Pelayo (1983). This is a weighted least-squares
iterative procedure, including terms in the eccentricity up to the fifth order. There are five
independent variables $(T_0, P_s, e, \dot{\omega}, \omega_0)$ determined in this procedure. The
periastron position $\omega$ is given by the linear equation

\medskip
\noindent $ \omega = \omega_0 + \dot{\omega}\ E $,

\medskip
\noindent where $\dot{\omega}$ is the rate of periastron advance, and the
position of periastron for the zero epoch $T_0$ is denoted as $\omega_0$.
The relation between the sidereal and the anomalistic period,
$P_s$ and $P_a$, is given by

\medskip
\noindent $ P_s = P_a \,(1 - \dot{\omega}/360^\circ), $

\medskip
\noindent and the period of apsidal motion by

\medskip
\noindent $ U = 360^\circ P_a/\dot{\omega} $.

\medskip
\noindent
All new precise CCD times of minima were used with a weight of 10
or 20 in our computation.
Some of our less precise measurements were weighted by a factor of 5,
while the earlier visual and photographic times (esp. the times of
the mid-exposure of a photographic plate) were given a weight of one
or nought because of the large scatter in these data.


\section{Analysis of the systems}

\subsection{HV 982}

The detached eclipsing binary \object{HV 982} (also known as MACHO 82.8043.26, LMV110; $V_{\rm
max}$ = 14\m65; Sp. B5V) is a relatively well-known LMC binary with an eccentric orbit ($e = 0.15$)
and a moderate orbital period of 5.3 days. It was discovered to be a variable star by Gaposhkin
(1970), who published the first photographic light curve.

Pritchard et al. (1994, 1998) in their photometric study derived high surface temperatures ($\sim$
28 000 K) and masses ($\sim$ 8 \ms) of components. They also derived the apsidal motion period $U =
205 \pm 7$ years. The precise stellar parameters of components of \object{HV~982} were derived
spectroscopically by Fitzpatrick et al. (2002), who found components with similar mass and size,

\begin{figure}
\includegraphics[width=0.5\textwidth]{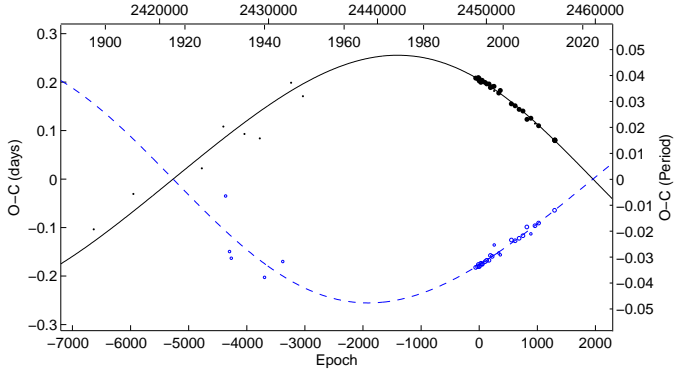}
\caption[ ]{$O-C$\ diagram for the times of minima of HV~982.
    The continuous and dashed curves represent predictions for the primary
    and secondary eclipses, respectively. The individual primary and
    secondary minima are denoted by dots and open circles, respectively.
    Larger symbols correspond to the photoelectric or CCD measurements
    that were given higher weights in the calculations.}
\label{hv982}
\end{figure}

\begin{center}
            $M_1 = 11.28 \pm 0.46$ \ms, $M_2 = 11.61 \pm 0.47$ \ms, \\
            $R_1 = 7.15 \pm 0.12$ \rs, $R_2 = 7.92 \pm 0.13$ \rs.
\end{center}

\noindent
The following linear light elements were given in that paper:

\begin{center}
Pri. Min. = HJD 24 49340$\fd$7172 + 5$\fd$33522 $\times$ E.
\end{center}

\noindent Using the complete analyses of HV~982 they also found the distance to the center of the
LMC $d_{LMC} = 50.7 \pm 1.2$ kpc. Later, Clausen et al. (2003) presented a new accurate CCD $uvby$
light curve obtained at the Danish 1.54 m telescope at La Silla and derived the precise photometric
elements with apsidal motion period $U = 208 \pm 15$ years.

All CCD times of minimum light given in Pritchard et al. (1998) and Clausen et al. (2003) were
incorporated into our analysis. Using {\sc Macho} (Faccioli et al. 2007) and {\sc Ogle} (Graczyk et
al. 2011) photometry, we were able to derive additional times of minimum light. A total of 68 times
of minimum light were used in our analysis (see Table~\ref{minima}). The orbital inclination
adopted was $i = 89\fdg3$, based on the analysis of Fitzpatrick et al. (2002). The computed apsidal
motion parameters and their internal errors of the least-squares fit are given in Table~\ref{t2}.
In this table, $P_s$ denotes the sidereal period, $P_a$ the anomalistic period, $e$ represents the
eccentricity, and $\dot{\omega}$ is the rate of periastron advance (in degrees per cycle and in
degrees per year). The zero epoch is given by $T_0$, and the corresponding position of the
periastron is represented by $\omega_0$. The $O-C$\ residuals for all times of minimum with respect
to the linear part of the apsidal motion equation ($\mathrm{Min} = \mathrm{T_0} + \mathrm{P} \times
\mathrm{E}$) are shown in Fig.~\ref{hv982}. The non-linear predictions, corresponding to the fitted
parameters, are plotted for primary and secondary eclipses.

\begin{table*}
\caption{Apsidal motion elements for
                 HV~982, HV~2274, MACHO~78.6097.13, MACHO~81.8881.47, and MACHO~79.5377.76.}
\label{t2}
\begin{flushleft}
\begin{tabular}{lcccccc}
\hline\hline\noalign{\smallskip}
Element [Unit]                          &   HV~982          &   HV~2274        & MACHO~78.6097.13 &  MACHO~81.8881.47  &  MACHO~79.5377.76 \\
\noalign{\smallskip}\hline\noalign{\smallskip}
$T_0$ [HJD]                             &24 49340.5146 (195)& 24 48099.8540(9) &24 52424.6888 (78)& 24 52282.2431 (178)& 24 52262.6776 (67)\\
$P_s$ [days]                            & 5.3352210 (97)    & 5.7259971 (8)    & 3.1070278 (104)  & 3.8818717 (32)     & 2.6365767 (79)  \\
$P_a$ [days]                            & 5.3355902 (97)    & 5.7267045 (8)    & 3.1075678 (104)  & 3.8822732 (33)     & 2.6370342 (79)  \\
$e$                                     & 0.149 (0.033)     & 0.1252 (39)      & 0.0459 (139)     & 0.217 (23)         & 0.0574 (160)    \\
$\dot{\omega}$ [deg $\:\rm{cycle^{-1}}$]& 0.0249 (0.0014)   & 0.0445 (29)      & 0.0633 (264)     & 0.03723 (606)      & 0.0625 (194)   \\
$\dot{\omega}$ [deg $\:\rm{yr^{-1}}$]   & 1.705 (0.095)     & 2.84 (0.20)      & 7.44 (3.1)       & 3.50 (0.57)        & 8.65 (2.68)   \\
$\omega_0$ [deg]                        & 221.2 (0.7)       & 71.9 (2.9)       & 27.9 (4.0)       & 95.9 (8.6)         & 31.57 (4.20) \\
$U$ [yr]                                & 211 (12)          & 126.9 (7.9)      & 48.4 (12.5)      & 102.8 (20.0)       & 41.6 (18.8) \\
 \noalign{\smallskip}\hline
\end{tabular}
\end{flushleft}
\end{table*}


\subsection{HV 2274}

\begin{figure}
\includegraphics[width=0.5\textwidth]{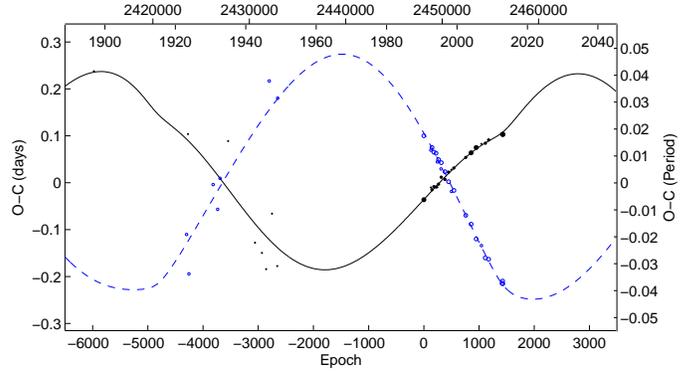}
\caption[]{$O-C$\ diagram for the times of minima of HV~2274. See legend to Fig.~\ref{hv982}. The
final fit is composed from two effects: the apsidal motion together with the third-body
hypothesis.} \label{hv2274}
\end{figure}

\begin{figure}
\includegraphics[width=0.5\textwidth]{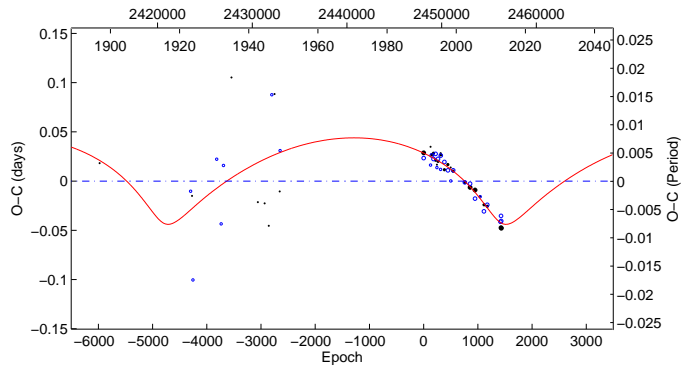}
\caption[]{$O-C$\ residuals for HV~2274 after subtraction of the apsidal motion term. Only the
third body effect is plotted.} \label{hv2274LITE}
\end{figure}

The detached and double-lined eclipsing binary \object{HV~2274} (also known as MACHO~19.3577.7,
2MASS~J05024076-6824212, LMV~182, FL~3556; $V_{\rm max}$ = 14\m13; Sp. B1-2 IV-III) is a relatively
bright and well-studied LMC eclipsing binary with an eccentric orbit ($e = 0.17$) and a moderate
orbital period of 5.7 days. It was discovered to be a variable star by Ms. Leavitt (1908). Later,
Shapley \& Nail (1953) recognized its eclipsing nature and classified it as a $\beta$~Lyrae type.
The eccentric orbit and apsidal motion of \object{HV~2274} was first announced by Watson et al.
(1992), who obtained the $BVI_c$ CCD photometry at the Mount John University Observatory. See also
the history of \object{HV~2274} in the last paper mentioned. Later Claret (1996) found the masses
and evolutionary status of this system:

\begin{center}
Pri. Min. = HJD 24 48099$\fd$818 + 5$\fd$726006 $\times$ E.
\end{center}

\noindent The fundamental properties of HV~2274 were most recently given in Ribas et al. (2000) who
determined the precise absolute parameters of both eclipsing components to be

\begin{center}
              $M_1 = 12.2 \pm 0.7$ \ms, $M_2 = 11.4 \pm 0.7$ \ms, \\
              $R_1 = 9.86 \pm 0.24$ \rs, $R_2 = 9.03 \pm 0.24$ \rs.
\end{center}

\begin{table}[b]
\caption{Third-body orbit parameters for HV~2274.} \label{tLITE}
\begin{flushleft}
\begin{tabular}{lcccccc}
\hline\hline\noalign{\smallskip}
Element [Unit]    &   Value \\
\noalign{\smallskip}\hline\noalign{\smallskip}
 $p_3$ [yr]        & 98.2 $\pm$ 14.3  \\
 $A_3$ [day]       & 0.045 $\pm$ 0.009 \\
 $T_3$ [HJD]       &2456205 $\pm$ 5033 \\
 $e_3$             & 0.654 $\pm$ 0.047 \\
 $\omega_3$ [deg]  & 251.1 $\pm$ 7.1   \\
 \noalign{\smallskip}\hline
\end{tabular}
\end{flushleft}
\end{table}

\noindent They also derived the improved apsidal motion period $U = 123\pm3$ years and the value of
internal structure constant log $k_{2,obs} = -2.56$. Since the above-mentioned papers were
published, new times of minima have been obtained, which allowed us to reduce the uncertainties in
the derived parameters. We collected all times of minimum light given in the literature together
with new ones derived from {\sc Macho}, {\sc Ogle}, and our new photometry obtained in Chile. All
of these values are listed in Table~\ref{minima}. In total, 58 precise times of minimum light were
used in our analysis, including 29 secondary eclipses. The orbital inclination adopted was $i =
89\fdg6$, based on the analysis of Ribas et al. (2000).

Analysing the available data using the apsidal motion hypothesis, we found an additional variation
superimposed on the apsidal motion. Hence, we used a different code computing the apsidal motion
parameters together with the third-body orbit (a so-called light travel time effect), see e.g.
Irwin (1959). Altogether, ten parameters were fitted (five from apsidal motion, five from the third
body hypothesis), thus this approach led to an acceptable solution. The resulting parameters of the
fit are given in Tables~\ref{t2} and \ref{tLITE}, the complete $O-C$\ diagrams are shown in
Figs.~\ref{hv2274} and \ref{hv2274LITE}. From the third-body parameters we were also able to
compute the mass function of the distant component, which resulted in $f(m_3) = 0.053 \pm 0.008$
\ms. From this value, we calculated the predicted minimal mass of the third body (i.e. assuming
coplanar orbits $i_3=90^\circ$), which resulted in $m_{3,min} = 3.4$ \ms. If we propose this body
in the system, one can ask whether it is detectable somehow in the already-obtained data. The
period is rather long for continuous monitoring of the radial velocity changes, but detecting the
third light in the light curve solution is difficult. Assuming a normal main sequence star, its
luminosity is about only 1\% of the total system luminosity. Such a weak third light could be
detectable only in extremely precise photometric data for the light curve solution.


\subsection{MACHO 78.6097.13}

\begin{figure}
\includegraphics[width=0.5\textwidth]{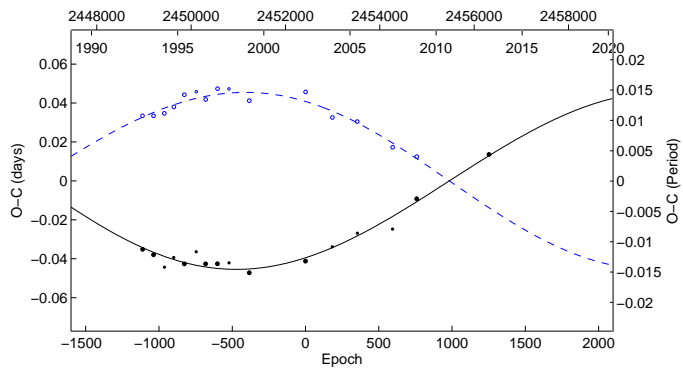}
\caption[]{$O-C$\ graph for the times of minimum of MACHO~78.6097.13 .
           See legend to Fig.~\ref{hv982}.}
\label{lmc5}
\end{figure}

The detached eclipsing binary \object{MACHO 78.6097.13} (also known as OGLE~J051804.81-694818.9,
LMC~MP\#5; $V_{\rm max}$ = 14\m37, sp O9V+O9V) is an eccentric binary system ($e=0.046$) with a
short orbital period ($P=3.1$ d) discovered by MP05 in the {\sc Ogle} field LMC-SC7.

\noindent Our analysis gives the following ephemeris:

\begin{center}
 Pri. Min. = HJD 24 52424$\fd$6888 + 3$\fd$1070278 $\times$ E.
\end{center}

\noindent The most detailed analysis of the system was published by Gonz{\'a}lez et al. (2005),
which was based on the ESO spectral observations together with the MACHO data. They also derived
the masses and radii of both components that were used in the present analysis. Both components are
of O9 spectral types, which makes this system the earliest in our sample.

Our new times of minimum light, as well as the timings derived from {\sc Macho} and {\sc Ogle}
photometry are given in Table~\ref{minima}. All of these data (31 minima times) were used in our
calculations. Using the parameters presented by Gonz{\'a}lez et al. (2005), we analysed the system
on apsidal motion, deriving the parameters given in Table \ref{t2}. As one can see, this system has
the lowest eccentricity in our sample and the apsidal advance is rather fast: almost one half of
the period has been covered with observations so far.


\begin{table}
\caption{Light and radial velocity curve fit parameters for MACHO~81.8881.47 and MACHO~79.5377.76.}
\label{tabLC}
\begin{flushleft}
\begin{tabular}{lcccccc}
\hline\hline\noalign{\smallskip}
Parameter        & MACHO~81.8881.47 & MACHO~79.5377.76\\
 \noalign{\smallskip}\hline\noalign{\smallskip}
  $T_1$ [K]      & 17200 (fixed)    & 27500 (fixed) \\
  $T_2$ [K]      & 18420 (360)      & 26710 (470)   \\
  $i$ [deg]      & 84.20 (0.27)     & 87.22 (0.44)  \\
  $\Omega_1$     & 7.067 (0.115)    & 5.774 (0.085) \\
  $\Omega_2$     & 6.365 (0.102)    & 6.264 (0.098) \\
  $q=M_2/M_1$    & 0.981 (0.030)    & 1.00 (0.02)   \\
  $a$ [R$_\odot$]& 23.04 (0.12)     & 22.65 (0.07)  \\
  $v_\gamma$ [km/s]& 272.6 (1.6)      & 250.0 (1.0)   \\
  $L_1$ [\%]     & 41.4 (1.2)       & 57.2 (1.5)    \\
  $L_2$ [\%]     & 58.6 (1.4)       & 42.8 (1.1)    \\
 \noalign{\smallskip}\hline
\end{tabular}
\end{flushleft}
\end{table}

\subsection{MACHO~81.8881.47}

\begin{figure}
 \includegraphics[width=0.5\textwidth]{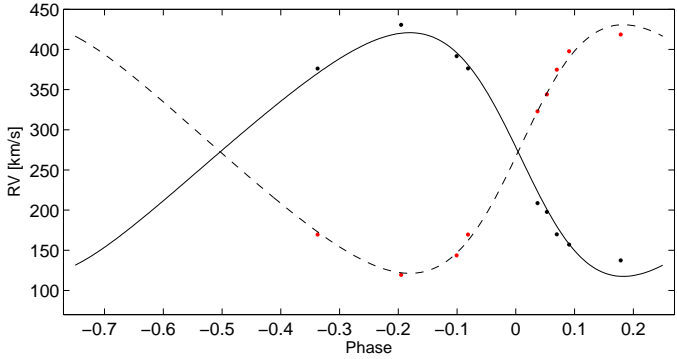}
 \caption[]{Radial velocity curve for MACHO~81.8881.47.} \label{LMC07RV}
\end{figure}

\begin{figure}
 \includegraphics[width=0.5\textwidth]{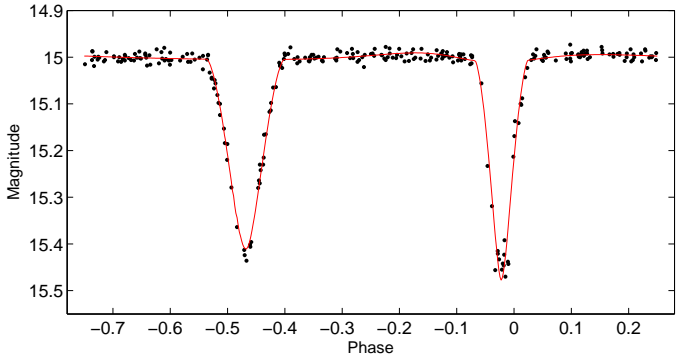}
 \caption[]{Light curve for MACHO~81.8881.47.} \label{LMC07LC}
\end{figure}

\begin{figure}
\includegraphics[width=0.5\textwidth]{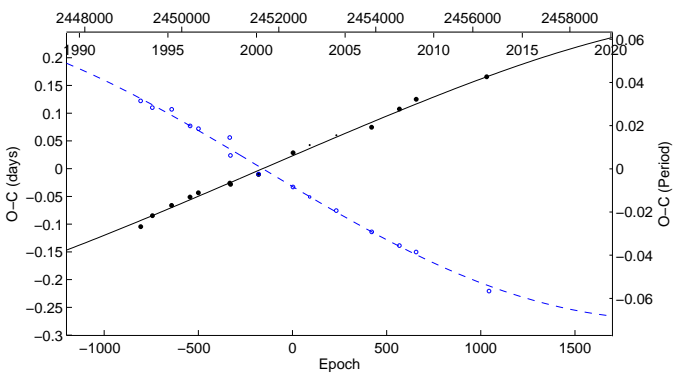}
\caption[]{$O-C$\ graph of MACHO~81.8881.47.
           See legend for Fig.~\ref{hv982}.}
\label{lmc7}
\end{figure}

The detached eclipsing binary \object{MACHO~81.8881.47} (also known as OGLE J053517.75-694318.7,
LMC MP\#7; $V_{\rm max}$ = 14\m9; Sp. B) is a relatively bright binary system with an eccentric
orbit ($e=0.2$) and a moderate orbital period $P \simeq 3.9$ days. Its variability was discovered
by MP05 in the {\sc OGLE} field LMC-SC16. Most recently, Graczyk et al. (2011) included this star
in their catalogue of eclipsing binaries in the LMC. They also gave the preliminary ephemeris:

\begin{center}
Pri. Min. = HJD 24 53571$\fd$1063 + 3$\fd$881980 $\times$ E.
\end{center}

\noindent Using {\sc Ogle} and {\sc Macho} photometry, we were also able to derive additional times
of minimum light; two more minima were derived from our observations. Only 30 times of minimum
light were used in our analysis (see Table~\ref{minima}). As one can clearly see from the shape of
the light curve and the different duration of both the primary and secondary eclipse, this is the
system with the highest value of eccentricity in our sample.

We used the ESO data for deriving the radial velocities of the components. Ten {\sc UVES} spectra
were found, of which nine were usable for the analysis. The derived RVs of both components are
given in Table \ref{RVs} in the Appendix. Together with these radial velocities, the MACHO light
curve was used for the subsequent combined LC+RV analysis. The final parameters of the solution are
given in Table \ref{tabLC}, while the fits are plotted in Figs. \ref{LMC07RV} and \ref{LMC07LC}.
Our results show the rather noticeable property that the more massive component (the primary) is
smaller and less luminous (i.e. has a lower temperature). However, this is still a preliminary
result based on only nine RVs and a poor fit. Although the scenario is possible, the more probable
explanation is that the mass ratio is inverse, and the primary and secondary components are
interchanged. This can be allowed for within the uncertainties of the mass ratio. A more detailed
analysis is needed, based on more precise spectral observations. The orbital inclination was about
$i = 84\fdg2$, which was later used for the apsidal motion analysis. The resulting parameters of
apsidal motion are found in Table~\ref{t2}, and the current $O-C$\ diagram is shown in
Fig.~\ref{lmc7}. As one can see, the period is rather long and only about 1/5 has been covered so
far.


\begin{figure}
 \includegraphics[width=0.5\textwidth]{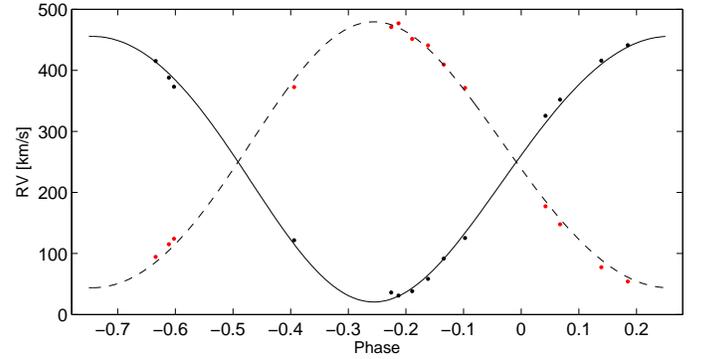}
 \caption[]{Radial velocity curve for MACHO~79.5377.76.} \label{OGLE10288RV}
\end{figure}

\begin{figure}
 \includegraphics[width=0.5\textwidth]{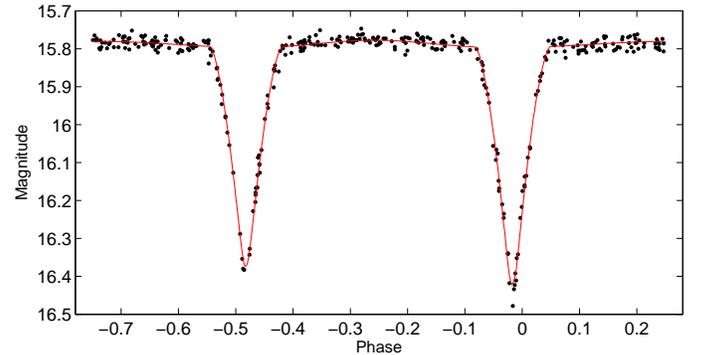}
 \caption[]{Light curve for MACHO~79.5377.76.} \label{OGLE10288LC}
\end{figure}

\begin{figure}
\includegraphics[width=0.5\textwidth]{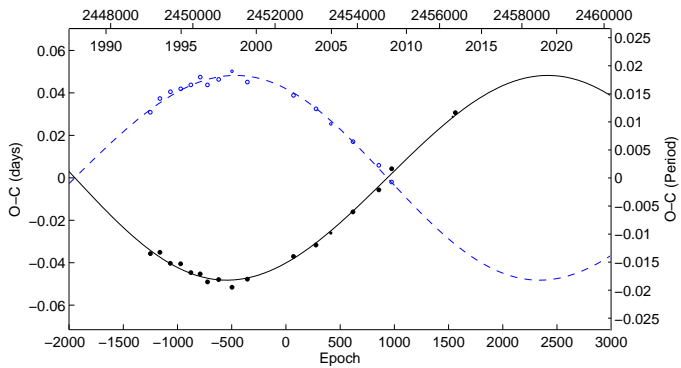}
\caption[]{$O-C$\ diagram for MACHO~79.5377.76. See legend to Fig.~\ref{hv982}.}
 \label{OGLE10288_OC}
\end{figure}
\subsection{MACHO~79.5377.76}

The detached eclipsing binary \object{MACHO~79.5377.76} (also known as OGLE J051323.98-692249.2,
$V_{\rm max}$ = 15\m8; Sp. B) is a fairly neglected binary system with a moderate eccentric orbit
($e=0.06$) and a short orbital period ($P \simeq 2.64$ day):

\begin{center}
Pri. Min. = HJD 24 52262$\fd$6776  + 2$\fd$6365767  $\times$ E.
\end{center}

We used the {\sc MACHO} and {\sc OGLE} photometry together with our new observations from Chile to
derive the times of minima, the ESO spectral observations to derive the radial velocities, and the
MACHO photometry to model the light curve of the system. Our results are plotted in Figs.
\ref{OGLE10288RV} and \ref{OGLE10288LC}. The resulting parameters are given in Table \ref{tabLC}.
As one can see, both components are of equal mass and their spectral type was estimated to be about
B0-B1. Radial velocities used for our analysis are also given in the Appendix in Table \ref{RVs}.
The coverage of the radial velocity curve is better than for MACHO~81.8881.47; however, it was
rather difficult to derive the synchronicity parameters $F_i$, and so we fixed them at $F_i = 0$. A
total of 34 reliable times of minimum light were used in our analysis including 16 secondary
eclipses (see tables in the Appendix). The final apsidal motion elements are given in
Table~\ref{t2}, and the $O-C$\ graph is shown in Fig.~\ref{OGLE10288_OC}. The apsidal period is the
shortest among systems studied here, only 42~years.


\section{Discussion}

\begin{table*}
\caption{Basic physical properties of HV~982, HV~2284, MACHO~78.6097.13,
 MACHO~81.8881.47, and MACHO~79.5377.76 and their internal structure constants}
\label{t3}
\begin{minipage}[]{\textwidth}
\begin{tabular}{lcccccccc}
\hline\hline\noalign{\smallskip}
Parameter & Unit                                     &  HV~982       & HV~2274     & MACHO~78.6097.13  & MACHO~81.8881.47 & MACHO~79.5377.76 \\
\hline\noalign{\smallskip}
$M_1$     & \ms                                      &  11.28 (0.46) & 12.2 (0.7)  &  11.76 (0.48)     & 5.51 (0.21)     &  11.26 (0.35)  \\
$M_2$     & \ms                                      &  11.61 (0.47) & 11.4 (0.7)  &  10.51 (0.40)     & 5.40 (0.19)     &  11.27 (0.35)  \\
$r_1$     &                                          &  0.194 (0.003)&0.255 (0.013)&  0.258 (0.021)    & 0.169 (0.008)   & 0.212 (0.004)\\
$r_2$     &                                          &  0.214 (0.003)&0.234 (0.012)&  0.206 (0.020)    & 0.192 (0.010)   & 0.190 (0.004)\\
\hline\noalign{\smallskip}
Source    &                                          & Clausen et al.& Ribas et al. &Gonz{\'a}lez et al.& This           & This    \\
          &                                          & (2003)        & (2000)       & (2005)            & paper          & paper   \\
\hline\noalign{\smallskip}
$\dot{\omega}_\mathrm{rel}$ & deg $\:\rm{cycle^{-1}}$& 0.0015        & 0.0014       & 0.0020            & 0.0011         & 0.0023  \\
$\dot{\omega}_\mathrm{rel} / \dot{\omega}$ & \%      & 5.9           & 3.2          & 3.24              & 3.06           & 3.65    \\
log $k_\mathrm{2, obs}$  &                           & -2.371 (0.10) &-2.470 (0.15) & -2.174 (0.35)     & -2.017 (0.18)  & -1.859 (0.17)\\
log $k_\mathrm{2, theo}$ &                           & -2.32         & -2.45        & -2.19             & -2.01          & -1.90   \\
\hline\noalign{\smallskip}
\end{tabular}
\end{minipage}
\end{table*}

The detection of apsidal motion in EEB provides the opportunity to test models of stellar internal
structure. The internal structure constant (ISC) $k_{2,\rm obs}$ is related to the variation in the
density inside the star and can be derived using the expression

\begin{equation}
k_\mathrm{2, obs} = \frac{1}{c_{21} + c_{22}} \, \frac{P_a}{U}
         = \frac{1}{c_{21} + c_{22}} \, \frac{\dot{\omega}}{360} ,
\end{equation}

\noindent where $c_{21}$ and $c_{22}$ are functions of the orbital eccentricity, fractional radii,
the masses of the components, and the ratio between rotational velocity of the stars and Keplerian
velocity (Kopal 1978). We also assume that the component stars rotate pseudosynchronously with the
same angular velocity as the maximum orbital value at periastron (see e.g. Kopal 1978). Another
possible approach is to use the value of $v\!\cdot\!\sin i$ as derived from the combined LC+RV
analysis published earlier. However, there could be a problem with the inclination of the rotation
axis (as in the case of DI~Her) and, moreover, the error of the internal structure constants is by
far dominated by the term $r_i^5$ in the equations. In addition to the classical Newtonian
contribution, the observed rate of rotation of the apses includes the contribution from General
Relativity (Gim\'enez 1985),

\begin{equation}
\dot{\omega}_\mathrm{rel} = \; 5.45 \times 10^{-4} \: \frac{1}{1-e^2}
          \:\Biggl( \frac{M_1 + M_2}{P} \Biggr) ^{2/3},
\end{equation}
\medskip

\noindent where $M_i$ denotes the individual masses of the
components in solar units and $P$ is the orbital period in days.

The values of $\dot{\omega}_\mathrm{rel}$ and the resulting mean internal structure constants
$k_\mathrm{2,obs}$ for the systems studied are given in Table~\ref{t3}. Theoretical values
$k_\mathrm{2,theo}$ according to available theoretical models for the internal stellar structure
computed by Claret (2006) for given masses of components are presented in Table \ref{t3}. 

\begin{figure}
\includegraphics[width=0.5\textwidth]{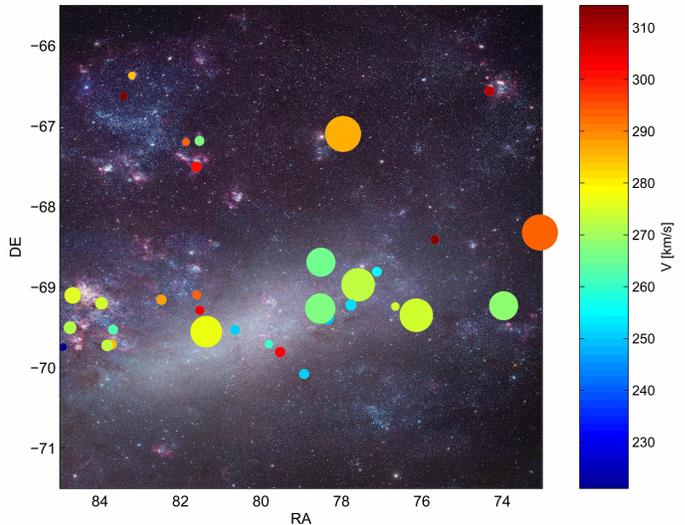}
\caption[]{Position and systemic velocities (see the colour scale on the right) of eclipsing
binaries located in the LMC and its vicinity, see Table \ref{TabRV_LMC}. The larger the symbol, the
higher the precision. The background image of the LMC is used with the permission of the author
Robert Gendler (http://www.robgendlerastropics.com).} \label{LMC_EBs}
\end{figure}

We tried to compare our resulting systemic velocities with other published values of eclipsing
binaries in the LMC, see Table \ref{TabRV_LMC} and Fig. \ref{LMC_EBs}. There are still only a few
such systems studied in detail (i.e. LC+RV analysis). Therefore, reliable analysis of different
velocities within the LMC is still very difficult. We can only compare our Fig. \ref{LMC_EBs} with
other kinematic studies of the LMC published earlier, e.g. that by  Reid \& Parker (2006) or Rohlfs
et al. (1984), that were based on much larger radial-velocity data sets. Nevertheless, we can
conclude that our two new systemic velocities roughly fit the overall picture and the total
velocity dispersion within the LMC is more than 30 km/s.

\begin{table*}
\caption{Radial velocities of eclipsing binaries in the LMC derived from various analyses.}
\label{TabRV_LMC}
\begin{minipage}[]{\textwidth}
\begin{tabular}{lcccccccc}
\hline\hline\noalign{\smallskip}
 System                & RA [h m s] & DE [$^\circ \,\, ^\prime \,\, ^{\prime\prime} $] & P [d] & RV [km/s] & Reference \\
  \hline\noalign{\smallskip}
 \object{MACHO 47.1884.17     } &  04 52 15.29   & -68 19 10.55   &251.0068   & 293.44$\pm$ 0.04 & Pietrzy{\'n}ski et al. (2013) \\
 \object{MACHO 18.2475.67     } &  04 55 51.48   & -69 13 47.99   &150.0198   & 267.68$\pm$ 0.08 & Pietrzy{\'n}ski et al. (2013) \\
 \object{HV 2241              } &  04 57 15.74   & -66 33 54.20   &  4.342635 & 307.9 $\pm$ 3    & Ostrov et al. (2001)          \\
         HV 2274                &  05 02 40.80   & -68 24 21.02   &  5.726006 & 312   $\pm$ 4    & Ribas et al. (2000)           \\
 \object{MACHO 1.3926.29      } &  05 04 32.88   & -69 20 50.99   &189.8215   & 274.32$\pm$ 0.05 & Pietrzy{\'n}ski et al. (2013) \\
 \object{MACHO 1.4290.113     } &  05 06 37.80   & -69 14 22.98   &  2.273210 & 274.2 $\pm$ 4.5  & Gonz{\'a}lez et al. (2005)    \\
 \object{MACHO 1.4539.37      } &  05 08 28.10   & -68 48 26.02   &  2.995450 & 255.6 $\pm$ 3.1  & Gonz{\'a}lez et al. (2005)    \\
 \object{OGLE LMC-ECL-9114    } &  05 10 19.65   & -68 58 12.0    &214.1707   & 272.04$\pm$ 0.05 & Pietrzy{\'n}ski et al. (2013) \\
 \object{MACHO 79.5017.83     } &  05 11 02.80   & -69 13 09.01   &  2.152915 & 252.0 $\pm$ 2.0  & Gonz{\'a}lez et al. (2005)    \\
 \object{MACHO 52.5169.24     } &  05 11 49.45   & -67 05 45.20   &167.6350   & 286.24$\pm$ 0.04 & Pietrzy{\'n}ski et al. (2013) \\
         MACHO 79.5377.76       &  05 13 23.91   & -69 22 48.90   &  2.636577 & 250.0 $\pm$ 1.0  & This paper                    \\
 \object{MACHO 2.5509.50      } &  05 14 01.91   & -68 41 18.41   &117.8708   & 265.10$\pm$ 0.08 & Pietrzy{\'n}ski et al. (2013) \\
 \object{MACHO 79.5500.60     } &  05 14 05.95   & -69 15 56.83   &771.7806   & 266.38$\pm$ 0.07 & Pietrzy{\'n}ski et al. (2013) \\
 \object{MACHO 6.5730.3092    } &  05 15 41.50   & -70 04 39.00   &  1.761014 & 250.9 $\pm$ 2.6  & Gonz{\'a}lez et al. (2005)   \\
         MACHO 78.6097.13       &  05 18 04.70   & -69 48 19.02   &  3.107023 & 301.7 $\pm$ 2.4  & Gonz{\'a}lez et al. (2005)   \\
 \object{HV 12012             } &  05 19 11.78   & -69 42 24.38   &  2.727125 & 261.4 $\pm$ 4.6  & Ribas et al. (2002)          \\
 \object{MACHO 78.6827.66     } &  05 22 35.00   & -69 31 44.01   &  2.183358 & 250.8 $\pm$ 3.0  & Gonz{\'a}lez et al. (2005)   \\
 \object{MACHO 77.7311.102    } &  05 25 25.55   & -69 33 04.49   &157.3243   & 276.66$\pm$ 0.06 & Pietrzy{\'n}ski et al. (2013)\\
 \object{MACHO 80.7436.52     } &  05 26 04.40   & -69 17 10.99   &  1.664135 & 303.2 $\pm$ 3.8  & Gonz{\'a}lez et al. (2005)  \\
 \object{TYC 8891-3349-1      } &  05 26 06.15   & -67 10 56.98   &  3.30161  & 267   $\pm$ 3    & Ostrov \& Lapasset (2003)   \\
 \object{MACHO 80.7438.42     } &  05 26 21.60   & -69 05 45.00   &  1.505947 & 292.7 $\pm$ 4.4  & Gonz{\'a}lez et al. (2005)  \\
 \object{$[$L72$]$ LH 54-425  } &  05 26 24.25   & -67 30 17.19   &  2.24741  & 300.9 $\pm$ 2.3  & Williams et al. (2008)     \\
 \object{HV 2543              } &  05 27 27.40   & -67 11 54.55   &  4.829046 & 293.2 $\pm$ 6    & Ostrov et al. (2000)      \\
         HV 982                 &  05 29 53.00   & -69 09 22.99   &  5.335220 & 287.8 $\pm$ 2.5  & Fitzpatrick et al. (2002)\\
 \object{LMC X-4              } &  05 32 49.54   & -66 22 13.30   &  1.40830  & 284.0 $\pm$ 7.0  & Hutchings et al. (1978)  \\
 \object{HV 5936              } &  05 33 39.03   & -66 37 39.61   &  2.805068 & 314.3 $\pm$ 5.8  & Fitzpatrick et al. (2003)\\
 \object{MACHO 81.8881.21     } &  05 34 48.14   & -69 42 36.30   &  4.250806 & 284   $\pm$ 3    & Bonanos (2009)          \\
 \object{MACHO 81.8763.8      } &  05 34 41.30   & -69 31 39.01   &  1.404740 & 263.1 $\pm$ 3    & Ostrov (2001)           \\
         MACHO 81.8881.47       &  05 35 17.57   & -69 43 18.90   &  3.881872 & 272.6 $\pm$ 1.6  & This paper              \\
 \object{MACHO 82.9010.36     } &  05 35 50.79   & -69 12 00.44   &  2.762456 & 273.8 $\pm$ 1.2  & Massey et al. (2012)    \\
 \object{$[$HSH95$]$ 38       } &  05 38 42.10   & -69 06 07.79   &  3.39     & 275.3 $\pm$ 0.5  & Massey et al. (2002)    \\
 \object{$[$HSH95$]$ 39       } &  05 38 42.49   & -69 06 01.29   &  4.06     & 266.9 $\pm$ 0.5  & Massey et al. (2002)    \\
 \object{$[$HSH95$]$ 42       } &  05 38 42.18   & -69 06 02.38   &  2.89     & 272.0 $\pm$ 0.5  & Massey et al. (2002)    \\
 \object{$[$HSH95$]$ 77       } &  05 38 42.56   & -69 06 04.39   &  1.88     & 275.2 $\pm$ 0.5  & Massey et al. (2002)    \\
 \object{LMC X-3              } &  05 38 56.63   & -64 05 03.30   &  1.7049   & 310.0 $\pm$ 7.0  & Cowley et al. (1983)    \\ 
 \object{$[$M2002$]$ LMC 172231}&  05 38 58.10   & -69 30 11.31   &  3.225414 & 271.5 $\pm$ 1.2  & Massey et al. (2012)    \\
 \object{LMC X-1              } &  05 39 38.84   & -69 44 35.70   &  4.2288   & 221.0 $\pm$ 6.0  & Hutchings et al. (1987) \\
\hline\noalign{\smallskip}
\end{tabular}
\end{minipage}
\end{table*}


\section{Conclusions}

The apsidal motion in EEB has been used for decades to test evolutionary stellar models. This study
provides accurate information on the apsidal motion rates of five main-sequence early-type binary
systems in the LMC: \object{HV~982}, \object{HV~2274}, \object{MACHO~78.6097.13}, \object{MACHO
81.8881.47}, and \object{MACHO~79.5377.76}. In our Galaxy there are known a few hundreds of apsidal
motion EEBs, however in the LMC there are still only a few dozen of these systems (Michalska \&
Pigulski 2005). Hence this study still presents an important contribution to the topic. The
relativistic effects are weak, being up to 6\% of the total apsidal motion rate. For the systems
MACHO~79.5377.76 and MACHO~81.8881.47, their light and radial velocity curves were analysed for the
first time yielding the stellar parameters of both components given in Table \ref{t3}. Moreover,
when the MACHO~79.5377.76 internal structure constant and the theoretical model are compared the
system appears to be very young ($\sim 2\cdot 10^6$~yr).

In spite of the considerable amount of observational data that has been collected for decades, the
absolute dimensions of massive binary components are only known with an accuracy of about 1-3\%
(e.g. Clausen 2004). More detailed study based on more precise radial velocities would be very
profitable to derive the physical properties of components with higher accuracy. The most promising
system for further detailed analysis seems to be HV~2274 because of the putative third component in
the system. We still know only a few such systems nowadays, see e.g. Bozkurt \& De{\v g}irmenci
(2007), while HV~2274 is the first to be discovered out of our Galaxy. Only precise spectral
observations and their disentangling should reveal its true nature.

\medskip
\begin{acknowledgements}

We do thank the {\sc MACHO} and {\sc OGLE} teams for making all of the observations easily and
publicly available. This work was supported by the Research Program MSM0021620860 {\it Physical
Study of objects and processes in the Solar System and in Astrophysics} of the Ministry of
Education of the Czech Republic, by the Czech Science Foundation grant no. P209/10/0715, by the
grant UNCE 12 of the Charles University in Prague, and by the grant LG12001 of the Ministry of
Education of the Czech Republic. We are also grateful to the ESO team at the La Silla Observatory
for their help in maintaining and operating the Danish telescope. The authors would like to thank
M.Zejda, J.Li\v{s}ka, J.Jan\'{\i}k, and M.Skarka for their important help with photometric
observations. G.Michalska is also acknowledged for sending us the unpublished times of minima for
selected binaries. The following internet-based resources were used in research for this paper: the
SIMBAD database and the VizieR service operated at the CDS, Strasbourg, France, the NASA
Astrophysics Data System Bibliographic Services, and the ESO Science Archive Facility.
\end{acknowledgements}



\begin{appendix} 

\section{Tables of minima}

\begin{table*}
\tiny \caption{List of the minima timings used for the analysis} \label{minima}
\begin{tabular}{lllrcl}
\hline\hline\noalign{\smallskip}
 Star       &    JD Hel.- &  Error & Epoch   &  Filter  & Source       \\
            &   2400000   &  [day] &         &          & Observatory  \\
\noalign{\smallskip}\hline
\noalign{\smallskip}
 HV~982 &  13946.555   &  0.05     &  -6634.0  &        & Gaposhkin (1977) \\
 HV~982 &  17590.584   &  0.05     &  -5951.0  &        & Gaposhkin (1977) \\
 HV~982 &  23875.527   &  0.05     &  -4773.0  &        & Gaposhkin (1977) \\
 HV~982 &  25849.645   &  0.05     &  -4403.0  &        & Gaposhkin (1977) \\
 HV~982 &  26060.243   &  0.10     &  -4363.5  &        & Gaposhkin (1977) \\
 HV~982 &  26412.253   &  0.05     &  -4297.5  &        & Gaposhkin (1977) \\
 HV~982 &  26577.631   &  0.05     &  -4266.5  &        & Gaposhkin (1977) \\
 HV~982 &  27786.315   &  0.05     &  -4040.0  &        & Gaposhkin (1977) \\
 HV~982 &  29189.469   &  0.10     &  -3777.0  &        & Gaposhkin (1977) \\
 HV~982 &  29629.338   &  0.05     &  -3694.5  &        & Gaposhkin (1977) \\
 HV~982 &  31304.630   &  0.05     &  -3380.5  &        & Gaposhkin (1977) \\
 HV~982 &  32070.603   &  0.05     &  -3237.0  &        & Gaposhkin (1977) \\
 HV~982 &  33153.625   &  0.05     &  -3034.0  &        & Gaposhkin (1977) \\
 HV~982 &  49335.3866  &  0.0004   &   -1.0    &$uvby$VI& Pritchard et al. (1998) \\
 HV~982 &  49337.6668  &  0.0004   &   -0.5    &$uvby$VI& Pritchard et al. (1998) \\
 HV~982 &  49337.6670  &  0.0010   &   -0.5    & $uvby$ & Clausen et al. (2003) \\
 HV~982 &  49340.7172  &  0.0005   &   0.0     & $uvby$ & Clausen et al. (2003) \\
 HV~982 &  50695.852   &  0.11     &   254.0   &$uvby$VI& Pritchard et al. (1998) \\
 HV~982 &  49186.00052 &  0.01299  &  -29.0    &   R    & MACHO, this paper \\
 HV~982 &  49188.28011 &  0.00421  &  -28.5    &   R    & MACHO, this paper \\
 HV~982 &  49618.14725 &  0.00256  &  52.0     &   R    & MACHO, this paper \\
 HV~982 &  49620.43922 &  0.01275  &  52.5     &   R    & MACHO, this paper \\
 HV~982 &  50023.61929 &  0.00560  &  128.0    &   R    & MACHO, this paper \\
 HV~982 &  50025.92309 &  0.00405  &  128.5    &   R    & MACHO, this paper \\
 HV~982 &  50370.40101 &  0.00352  &  193.0    &   R    & MACHO, this paper \\
 HV~982 &  50367.38712 &  0.00629  &  192.5    &   R    & MACHO, this paper \\
 HV~982 &  50711.84825 &  0.02668  &  257.0    &   R    & MACHO, this paper \\
 HV~982 &  50714.19804 &  0.02646  &  257.5    &   R    & MACHO, this paper \\
 HV~982 &  51277.38263 &  0.01014  &  363.0    &   R    & MACHO, this paper \\
 HV~982 &  51279.71120 &  0.03201  &  363.5    &   R    & MACHO, this paper \\
 HV~982 &  49020.60964 &  0.00588  &  -60.0    &   B    & MACHO, this paper \\
 HV~982 &  49017.55154 &  0.00615  &  -60.5    &   B    & MACHO, this paper \\
 HV~982 &  49260.69563 &  0.00240  &  -15.0    &   B    & MACHO, this paper \\
 HV~982 &  49262.97768 &  0.00197  &  -14.5    &   B    & MACHO, this paper \\
 HV~982 &  49490.10034 &  0.00331  &  28.0     &   B    & MACHO, this paper \\
 HV~982 &  49487.05952 &  0.00848  &  27.5     &   B    & MACHO, this paper \\
 HV~982 &  49612.81440 &  0.00315  &  51.0     &   B    & MACHO, this paper \\
 HV~982 &  49609.76750 &  0.00955  &  50.5     &   B    & MACHO, this paper \\
 HV~982 &  49879.57123 &  0.002614 &  101.0    &   B    & MACHO, this paper \\
 HV~982 &  49881.86935 &  0.00480  &  101.5    &   B    & MACHO, this paper \\
 HV~982 &  50231.69273 &  0.00384  &  167.0    &   B    & MACHO, this paper \\
 HV~982 &  50233.99656 &  0.004588 &  167.5    &   B    & MACHO, this paper \\
 HV~982 &  50530.45942 &  0.00416  &  223.0    &   B    & MACHO, this paper \\
 HV~982 &  50532.77709 &  0.00219  &  223.5    &   B    & MACHO, this paper \\
 HV~982 &  51133.32646 &  0.00331  &  336.0    &   B    & MACHO, this paper \\
 HV~982 &  51135.66413 &  0.00742  &  336.5    &   B    & MACHO, this paper \\
 HV~982 &  52293.43371 &  0.01771  &  553.5    &   I    & OGLE, this paper  \\
 HV~982 &  52291.04698 &  0.03068  &  553.0    &   I    & OGLE, this paper  \\
 HV~982 &  52632.49722 &  0.00320  &  617.0    &   I    & OGLE, this paper  \\
 HV~982 &  52634.88640 &  0.00213  &  617.5    &   I    & OGLE, this paper  \\
 HV~982 &  53005.95515 &  0.00261  &  687.0    &   I    & OGLE, this paper  \\
 HV~982 &  53008.35696 &  0.00726  &  687.5    &   I    & OGLE, this paper  \\
 HV~982 &  53355.15161 &  0.00636  &  752.5    &   I    & OGLE, this paper  \\
 HV~982 &  53358.07636 &  0.01552  &  753.0    &   I    & OGLE, this paper  \\
 HV~982 &  53720.85428 &  0.01008  &  821.0    &   I    & OGLE, this paper  \\
 HV~982 &  53717.96464 &  0.02203  &  820.5    &   I    & OGLE, this paper  \\
 HV~982 &  54083.65165 &  0.01595  &  889.0    &   I    & OGLE, this paper  \\
 HV~982 &  54086.08062 &  0.03     &  889.5    &   I    & OGLE, this paper  \\
 HV~982 &  54462.44122 &  0.02     &  960.0    &   I    & OGLE, this paper  \\
 HV~982 &  54464.89826 &  0.00320  &  960.5    &   I    & OGLE, this paper  \\
 HV~982 &  54793.22040 &  0.00373  &  1022.0   &   I    & OGLE, this paper  \\
 HV~982 &  54790.35208 &  0.00774  &  1021.5   &   I    & OGLE, this paper  \\
 HV~982 &  56257.56416 &  0.0001   & 1296.5    &   I    & DK154, this paper \\
 HV~982 &  56257.56421 &  0.0001   & 1296.5    &   I    & DK154, this paper \\
 HV~982 &  56265.71152 &  0.0001   & 1298.0    &   I    & DK154, this paper \\
 HV~982 &  56265.71137 &  0.0001   & 1298.0    &   I    & DK154, this paper \\
 HV~982 &  56297.72209 &  0.0003   & 1304.0    &   I    & DK154, this paper \\
 HV~982 &  56297.72347 &  0.0003   & 1304.0    &   I    & DK154, this paper \\ \hline
 HV~2274&  13875.807   &           & -5977.0   &    & Gaposhkin (1977)  \\
 HV~2274&  23486.545   &           & -4298.5   &    & Gaposhkin (1977)  \\
 HV~2274&  23638.498   &           & -4272.0   &    & Gaposhkin (1977)  \\
 HV~2274&  23732.679   &           & -4255.5   &    & Gaposhkin (1977)  \\
 HV~2274&  26246.582   &           & -3816.5   &    & Gaposhkin (1977)  \\
 HV~2274&  26710.335   &           & -3735.5   &    & Gaposhkin (1977)  \\
 HV~2274&  26956.619   &           & -3692.5   &    & Gaposhkin (1977)  \\
 HV~2274&  27801.283   &           & -3545.0   &    & Gaposhkin (1977)  \\
 HV~2274&  30589.627   &           & -3058.0   &    & Gaposhkin (1977)  \\
 HV~2274&  31299.629   &           & -2934.0   &    & Gaposhkin (1977)  \\
 HV~2274&  31740.496   &           & -2857.0   &    & Gaposhkin (1977)  \\
 HV~2274&  32058.690   &           & -2801.5   &    & Gaposhkin (1977)  \\
  \noalign{\smallskip}\hline
\end{tabular}
\end{table*}

\begin{table*}
\tiny \caption{List of the minima timings used for the analysis (cont.).} 
\begin{tabular}{lllrcl}
\hline\hline\noalign{\smallskip}
 Star            &    JD Hel.-  &  Error & Epoch   &  Filter  & Source       \\
                 &   2400000    &  [day] &         &          & Observatory  \\
\noalign{\smallskip}\hline \noalign{\smallskip}
 HV~2274&  32347.570   &           & -2751.0   &    & Gaposhkin (1977)  \\
 HV~2274&  32891.428   &           & -2656.0   &    & Gaposhkin (1977)  \\
 HV~2274&  32940.457   &           & -2647.5   &    & Gaposhkin (1977)  \\
 HV~2274&  48099.818   &           &     0.0   &    & Watson et al. (1992)     \\ 
 HV~2274&  48102.817   &           &     0.5   &    & Watson et al. (1992)     \\ 
 HV~2274&  48827.0456  &           &   127.0   &    & MP05             \\
 HV~2274&  48829.988   &           &   127.5   &    & MP05             \\
 HV~2274&  48947.28675 &  0.01254  &   148.0   &  B & MACHO, this paper  \\
 HV~2274&  48950.23969 &  0.01483  &   148.5   &  B & MACHO, this paper  \\
 HV~2274&  49136.25131 &  0.00567  &   181.0   &  B & MACHO, this paper  \\
 HV~2274&  49139.18771 &  0.01070  &   181.5   &  B & MACHO, this paper  \\
 HV~2274&  49359.56433 &  0.00636  &   220.0   &  B & MACHO, this paper  \\
 HV~2274&  49362.49934 &  0.00401  &   220.5   &  B & MACHO, this paper  \\
 HV~2274&  49499.9047  &           &   244.5   &    & MP05             \\
 HV~2274&  49502.7137  &           &   245.0   &    & MP05             \\
 HV~2274&  49614.42969 &  0.00578  &   264.5   &  B & MACHO, this paper  \\
 HV~2274&  49617.24018 &  0.00664  &   265.0   &  B & MACHO, this paper  \\
 HV~2274&  49889.2576  &           &   312.5   &    & MP05             \\
 HV~2274&  49892.104   &           &   313.0   &    & MP05             \\
 HV~2274&  49906.44901 &  0.01254  &   315.5   &  B & MACHO, this paper  \\
 HV~2274&  49909.28085 &  0.01763  &   316.0   &  B & MACHO, this paper  \\
 HV~2274&  50287.19199 &  0.00452  &   382.0   &  B & MACHO, this paper  \\
 HV~2274&  50290.07136 &  0.00767  &   382.5   &  B & MACHO, this paper  \\
 HV~2274&  50659.39741 &  0.00515  &   447.0   &  B & MACHO, this paper  \\
 HV~2274&  50662.24002 &  0.00425  &   447.5   &  B & MACHO, this paper  \\
 HV~2274&  50957.1546  &           &   499.0   &    & MP05             \\
 HV~2274&  50959.971   &           &   499.5   &    & MP05             \\
 HV~2274&  51214.82801 &  0.00315  &   544.0   &  B & MACHO, this paper  \\
 HV~2274&  51217.64304 &  0.00527  &   544.5   &  B & MACHO, this paper  \\
 HV~2274&  52445.93977 &  0.00596  &   759.0   &  I & OGLE, this paper   \\
 HV~2274&  52448.67936 &  0.01065  &   759.5   &  I & OGLE, this paper   \\
 HV~2274&  52998.35630 &  0.00504  &   855.5   &  I & OGLE, this paper   \\
 HV~2274&  53001.37141 &  0.00431  &   856.0   &  I & OGLE, this paper   \\
 HV~2274&  53528.17437 &  0.00624  &   948.0   &  I & OGLE, this paper   \\
 HV~2274&  53530.84233 &  0.00424  &   948.5   &  I & OGLE, this paper   \\
 HV~2274&  54069.07211 &  0.01346  &  1042.5   &  I & OGLE, this paper   \\
 HV~2274&  54072.15199 &  0.01866  &  1043.0   &  I & OGLE, this paper   \\
 HV~2274&  54469.86569 &  0.02703  &  1112.5   &  I & OGLE, this paper   \\
 HV~2274&  54472.97310 &  0.01248  &  1113.0   &  I & OGLE, this paper   \\
 HV~2274&  54819.14866 &  0.00470  &  1173.5   &  I & OGLE, this paper   \\
 HV~2274&  54822.26597 &  0.00618  &  1174.0   &  I & OGLE, this paper   \\
 HV~2274&  56204.78877 &  0.00079  &  1415.5   &  R & DK154, this paper  \\
 HV~2274&  56267.77936 &  0.00267  &  1426.5   &  R & DK154, this paper  \\
 HV~2274&  56270.95500 &  0.00257  &  1427.0   &  R & DK154, this paper  \\
 HV~2274&  56299.58505 &  0.00231  &  1432.0   &  R & DK154, this paper  \\
 HV~2274&  56290.67754 &  0.00178  &  1430.5   &  R & DK154, this paper  \\ \hline
 MACHO~78.6097.13 & 48968.15382 & 0.00342  & -1112.5 & BR & MACHO, this paper \\
 MACHO~78.6097.13 & 48969.63869 & 0.00206  & -1112.0 & BR & MACHO, this paper \\
 MACHO~78.6097.13 & 49201.18083 & 0.00205  & -1037.5 & BR & MACHO, this paper \\
 MACHO~78.6097.13 & 49202.66301 & 0.00385  & -1037.0 & BR & MACHO, this paper \\
 MACHO~78.6097.13 & 49431.10223 & 0.00306  &  -963.5 & BR & MACHO, this paper \\
 MACHO~78.6097.13 & 49432.57665 & 0.00549  &  -963.0 & BR & MACHO, this paper \\
 MACHO~78.6097.13 & 49631.43145 & 0.00707  &  -899.0 & BR & MACHO, this paper \\
 MACHO~78.6097.13 & 49633.06235 & 0.00284  &  -898.5 & BR & MACHO, this paper \\
 MACHO~78.6097.13 & 49855.13418 & 0.00370  &  -827.0 & BR & MACHO, this paper \\
 MACHO~78.6097.13 & 49856.77458 & 0.00531  &  -826.5 & BR & MACHO, this paper \\
 MACHO~78.6097.13 & 50106.80963 & 0.00559  &  -746.0 & BR & MACHO, this paper \\
 MACHO~78.6097.13 & 50108.44539 & 0.00551  &  -745.5 & BR & MACHO, this paper \\
 MACHO~78.6097.13 & 50305.65322 & 0.00573  &  -682.0 & BR & MACHO, this paper \\
 MACHO~78.6097.13 & 50307.29116 & 0.00378  &  -681.5 & BR & MACHO, this paper \\
 MACHO~78.6097.13 & 50557.32252 & 0.00262  &  -601.0 & BR & MACHO, this paper \\
 MACHO~78.6097.13 & 50558.96600 & 0.00634  &  -600.5 & BR & MACHO, this paper \\
 MACHO~78.6097.13 & 50801.31406 & 0.00683  &  -522.5 & BR & MACHO, this paper \\
 MACHO~78.6097.13 & 50802.77817 & 0.00996  &  -522.0 & BR & MACHO, this paper \\
 MACHO~78.6097.13 & 51231.54291 & 0.00308  &  -384.0 & BR & MACHO, this paper \\
 MACHO~78.6097.13 & 51233.18484 & 0.00640  &  -383.5 & BR & MACHO, this paper \\
 MACHO~78.6097.13 & 52424.64757 & 0.00294  &     0.0 & I  & OGLE, this paper \\
 MACHO~78.6097.13 & 52426.28805 & 0.00233  &     0.5 & I  & OGLE, this paper \\
 MACHO~78.6097.13 & 52993.24107 & 0.00798  &   183.0 & I  & OGLE, this paper \\
 MACHO~78.6097.13 & 52994.86101 & 0.00267  &   183.5 & I  & OGLE, this paper \\
 MACHO~78.6097.13 & 53524.54977 & 0.00558  &   354.0 & I  & OGLE, this paper \\
 MACHO~78.6097.13 & 53523.05361 & 0.00298  &   353.5 & I  & OGLE, this paper \\
 MACHO~78.6097.13 & 54270.23860 & 0.00578  &   594.0 & I  & OGLE, this paper \\
 MACHO~78.6097.13 & 54271.83416 & 0.00167  &   594.5 & I  & OGLE, this paper \\
 MACHO~78.6097.13 & 54782.91373 & 0.00432  &   759.0 & I  & OGLE, this paper \\
 MACHO~78.6097.13 & 54784.48892 & 0.00370  &   759.5 & I  & OGLE, this paper \\
 MACHO~78.6097.13 & 56314.70125 & 0.00179  &  1252.0 & R  & DK154, this paper\\ \hline
 MACHO~81.8881.47 & 49151.63566  & 0.0005  &    -0.5 &    & MP05             \\
 MACHO~81.8881.47 & 49153.34993  & 0.0005  &     0.0 &    & MP05             \\
  \noalign{\smallskip}\hline
\end{tabular}
\end{table*}

\begin{table*}
\tiny \caption{List of the minima timings used for the analysis (cont.).} 
\begin{tabular}{lllrcl}
\hline\hline\noalign{\smallskip}
 Star            &    JD Hel.-  &  Error & Epoch   &  Filter  & Source       \\
                 &   2400000    &  [day] &         &          & Observatory  \\
\noalign{\smallskip}\hline \noalign{\smallskip}
 MACHO~81.8881.47 & 49392.29958  & 0.0005  &    61.5 &    & MP05             \\
 MACHO~81.8881.47 & 49394.04567  & 0.0005  &    62.0 &    & MP05             \\
 MACHO~81.8881.47 & 49790.01510  & 0.0005  &   164.0 &    & MP05             \\
 MACHO~81.8881.47 & 49792.12920  & 0.0005  &   164.5 &    & MP05             \\
 MACHO~81.8881.47 & 50170.45366  & 0.0005  &   262.0 &    & MP05             \\
 MACHO~81.8881.47 & 50172.52273  & 0.0005  &   262.5 &    & MP05             \\
 MACHO~81.8881.47 & 50341.26363  & 0.0005  &   306.0 &    & MP05             \\
 MACHO~81.8881.47 & 50343.32028  & 0.0005  &   306.5 &    & MP05             \\
 MACHO~81.8881.47 & 50987.69498  & 0.0005  &   472.5 &    & MP05             \\
 MACHO~81.8881.47 & 50989.55366  & 0.0005  &   473.0 &    & MP05             \\
 MACHO~81.8881.47 & 51003.19013  & 0.0005  &   476.5 &    & MP05             \\
 MACHO~81.8881.47 & 51005.07908  & 0.0005  &   477.0 &    & MP05             \\
 MACHO~81.8881.47 & 51581.55557  & 0.0005  &   625.5 &    & MP05             \\
 MACHO~81.8881.47 & 51583.49576  & 0.0005  &   626.0 &    & MP05             \\
 MACHO~81.8881.47 & 52291.91465  & 0.0091  &   808.5 & I  & OGLE, this paper \\
 MACHO~81.8881.47 & 52293.91729  & 0.0037  &   809.0 & I  & OGLE, this paper \\
 MACHO~81.8881.47 & 52635.53595  & 0.0052  &   897.0 & I  & OGLE, this paper \\
 MACHO~81.8881.47 & 52637.38335  & 0.0877  &   897.5 & I  & OGLE, this paper \\
 MACHO~81.8881.47 & 53184.70263  & 0.0163  &  1038.5 & I  & OGLE, this paper \\
 MACHO~81.8881.47 & 53186.77939  & 0.1164  &  1039.0 & I  & OGLE, this paper \\
 MACHO~81.8881.47 & 53912.70392  & 0.0052  &  1226.0 & I  & OGLE, this paper \\
 MACHO~81.8881.47 & 53914.45626  & 0.0034  &  1226.5 & I  & OGLE, this paper \\
 MACHO~81.8881.47 & 54485.06658  & 0.0014  &  1373.5 & I  & OGLE, this paper \\
 MACHO~81.8881.47 & 54487.25375  & 0.0049  &  1374.0 & I  & OGLE, this paper \\
 MACHO~81.8881.47 & 54830.54160  & 0.0042  &  1462.5 & I  & OGLE, this paper \\
 MACHO~81.8881.47 & 54832.75781  & 0.0124  &  1463.0 & I  & OGLE, this paper \\
 MACHO~81.8881.47 & 56284.61939  & 0.0008  &  1837.0 & R  & DK154, this paper \\
 MACHO~81.8881.47 & 56332.75573  & 0.0010  &  1849.5 & R  & DK154, this paper \\ \hline
 MACHO~79.5377.76 & 48965.66934  & 0.00419 & -1250.5 & R  & MACHO, this paper \\ 
 MACHO~79.5377.76 & 48966.92100  & 0.00198 & -1250.0 & R  & MACHO, this paper \\ 
 MACHO~79.5377.76 & 49197.69453  & 0.00124 & -1162.5 & R  & MACHO, this paper \\ 
 MACHO~79.5377.76 & 49198.94037  & 0.00551 & -1162.0 & R  & MACHO, this paper \\ 
 MACHO~79.5377.76 & 49452.04656  & 0.00129 & -1066.0 & R  & MACHO, this paper \\ 
 MACHO~79.5377.76 & 49453.44570  & 0.00382 & -1065.5 & R  & MACHO, this paper \\ 
 MACHO~79.5377.76 & 49702.52111  & 0.00264 &  -971.0 & R  & MACHO, this paper \\ 
 MACHO~79.5377.76 & 49703.92184  & 0.00163 &  -970.5 & R  & MACHO, this paper \\ 
 MACHO~79.5377.76 & 49951.76191  & 0.00140 &  -876.5 & R  & MACHO, this paper \\ 
 MACHO~79.5377.76 & 49950.35515  & 0.00235 &  -877.0 & R  & MACHO, this paper \\ 
 MACHO~79.5377.76 & 50177.10012  & 0.00171 &  -791.0 & R  & MACHO, this paper \\ 
 MACHO~79.5377.76 & 50178.51120  & 0.00206 &  -790.5 & R  & MACHO, this paper \\ 
 MACHO~79.5377.76 & 50356.38361  & 0.00169 &  -723.0 & R  & MACHO, this paper \\ 
 MACHO~79.5377.76 & 50355.15816  & 0.00177 &  -723.5 & R  & MACHO, this paper \\ 
 MACHO~79.5377.76 & 50626.72815  & 0.00113 &  -620.5 & R  & MACHO, this paper \\ 
 MACHO~79.5377.76 & 50627.95207  & 0.00272 &  -620.0 & R  & MACHO, this paper \\ 
 MACHO~79.5377.76 & 50952.24739  & 0.00722 &  -497.0 & R  & MACHO, this paper \\ 
 MACHO~79.5377.76 & 50951.03095  & 0.04825 &  -497.5 & R  & MACHO, this paper \\ 
 MACHO~79.5377.76 & 51324.00854  & 0.00419 &  -356.0 & R  & MACHO, this paper \\ 
 MACHO~79.5377.76 & 51325.41971  & 0.00198 &  -355.5 & R  & MACHO, this paper \\ 
 MACHO~79.5377.76 & 52447.20093  & 0.00156 &    70.0 & I  & OGLE, this paper  \\
 MACHO~79.5377.76 & 52445.95855  & 0.00132 &    69.5 & I  & OGLE, this paper  \\
 MACHO~79.5377.76 & 52994.36017  & 0.00348 &   277.5 & I  & OGLE, this paper  \\
 MACHO~79.5377.76 & 52992.97769  & 0.00256 &   277.0 & I  & OGLE, this paper  \\
 MACHO~79.5377.76 & 53354.19440  & 0.00556 &   414.0 & I  & OGLE, this paper  \\
 MACHO~79.5377.76 & 53352.92752  & 0.01318 &   413.5 & I  & OGLE, this paper  \\
 MACHO~79.5377.76 & 53892.06597  & 0.00182 &   618.0 & I  & OGLE, this paper  \\
 MACHO~79.5377.76 & 53893.41737  & 0.00150 &   618.5 & I  & OGLE, this paper  \\
 MACHO~79.5377.76 & 54522.21822  & 0.00205 &   857.0 & I  & OGLE, this paper  \\
 MACHO~79.5377.76 & 54520.91151  & 0.00071 &   856.5 & I  & OGLE, this paper  \\
 MACHO~79.5377.76 & 54833.34419  & 0.00145 &   975.0 & I  & OGLE, this paper  \\
 MACHO~79.5377.76 & 54832.01969  & 0.00063 &   974.5 & I  & OGLE, this paper  \\
 MACHO~79.5377.76 & 56315.12485  & 0.005   &  1537.0 & R  & DK154, this paper \\
 MACHO~79.5377.76 & 56383.67777  & 0.00040 &  1563.0 & R  & DK154, this paper \\ \hline
\noalign{\smallskip}\hline
\end{tabular}
\end{table*}

\section{Tables of radial velocities}

\begin{table*}
\tiny \caption{List of the radial velocities used for the analysis.} \label{RVs}
\begin{tabular}{ccrr}
\hline\hline\noalign{\smallskip}
 Star             &  JD Hel.- & $RV_1$      & $RV_2$      \\
                  & 2400000   &[km s$^{-1}$]&[km s$^{-1}$]\\
\noalign{\smallskip}\hline \noalign{\smallskip}
 MACHO~81.8881.47 & 52243.62965 &  197.825  &  344.054  \\
 MACHO~81.8881.47 & 52243.69480 &  169.847  &  374.850  \\
 MACHO~81.8881.47 & 52243.77655 &  157.162  &  397.739  \\
 MACHO~81.8881.47 & 52246.54851 &  430.591  &  119.457  \\
 MACHO~81.8881.47 & 52257.64245 &  376.268  &  169.630  \\
 MACHO~81.8881.47 & 52270.74072 &  208.679  &  322.865  \\
 MACHO~81.8881.47 & 52289.69123 &  376.342  &  169.518  \\
 MACHO~81.8881.47 & 52294.58185 &  137.413  &  418.659  \\
 MACHO~81.8881.47 & 52320.67094 &  391.611  &  143.591  \\ \hline
 MACHO~79.5377.76 & 52214.79259 &   58.559  &  440.768  \\
 MACHO~79.5377.76 & 52226.72978 &  415.303  &   94.430  \\
 MACHO~79.5377.76 & 52243.72257 &   38.347  &  451.361  \\
 MACHO~79.5377.76 & 52244.58815 &  415.863  &   77.425  \\
 MACHO~79.5377.76 & 52244.70911 &  441.078  &   54.155  \\
 MACHO~79.5377.76 & 52245.81943 &  121.623  &  372.537  \\
 MACHO~79.5377.76 & 52246.60116 &  125.476  &  371.131  \\
 MACHO~79.5377.76 & 52270.69835 &  325.508  &  177.316  \\
 MACHO~79.5377.76 & 52270.76503 &  352.081  &  147.711  \\
 MACHO~79.5377.76 & 52271.61223 &  387.922  &  115.153  \\
 MACHO~79.5377.76 & 52271.63592 &  373.100  &  124.104  \\
 MACHO~79.5377.76 & 52296.59889 &   91.741  &  409.521  \\
 MACHO~79.5377.76 & 52309.54136 &   36.136  &  470.775  \\
 MACHO~79.5377.76 & 52309.57452 &   31.116  &  476.893  \\ \hline
\noalign{\smallskip}\hline
\end{tabular}
\end{table*}

\end{appendix}
\end{document}